\documentclass[utf8]{jetp}
\twocolumn 
\usepackage{amsmath}
\usepackage{graphicx}
\usepackage{amsfonts}
\usepackage{amssymb}

\begin{document}

\title{Non-local gravitational сorrections in black hole shadow images}

\setaffiliation1{Sternberg Astronomical Institute, Lomonosov Moscow State University, Universitetskii Prospekt, 13, Moscow, 119234, Russian Federation}

\setaffiliation2{Quantum theory and High Energy Physics Department, Physics Faculty, Lomonosov Moscow State University, Vorobievi Gori, 1/2, Moscow, 119234, Russian Federation}

\setaffiliation3{Ural Federal University named by First President of Russia B.N.Eltsin, Ulitsa Mira, 19, Yekaterinburg, 620002, Russian Federation}

\setauthor{S.}{Alexeyev}{12}
\setauthor{A.}{Baiderin}{2}
\setauthor{A.}{Nemtinova}{3}
\setauthor{O.}{Zenin}{2}

\date{December 4, 2023}

\abstract{With the help of Newman-Janis method new spinning black hole (BH) solution for a non-local gravity model was obtained. We show how to account the quantum gravitational correction part in BH shadows modelling using spinning BH metrics with a model independent approach. It is confirmed that in the future to follow the increasing of the experimental accuracy and therefore to reproduce new results theoretically one could take into account different field correction terms instead of introducing of new fields and/or curvature expansions.}

\maketitle

\section{Introduction}

The idea to use non-local actions in extended gravity models is discussed for a long time \cite{a01}. Using this approach gives hope to model the dark energy in a more natural way. Non-local constructions were used, for example, in Randall-Sundrum models \cite{a02}. Further considering of non-local additions allowed to set new constrains on gravity models using high energy physics data \cite{c03a}. So non-local operators also appear in the unique effective action for quantum gravity 
\begin{eqnarray}
L & = & R+c_1 R^2+ c_2 R_{\mu \nu}R^{\mu \nu}+c_3 R_{\mu\nu\alpha\beta}R^{\mu \nu  \alpha \beta} \nonumber \\ & + & \alpha R \log\frac{\Box}{\mu^2} R+ \beta
R_{\mu \nu} \log\frac{\Box}{\mu^2}  R^{\mu \nu} \nonumber \\ & + & \gamma R_{\mu \nu \alpha\beta} \log\frac{\Box}{\mu^2}  R^{\mu \nu  \alpha \beta}, \label{act00}
\end{eqnarray}
where $R$, $R_{\mu\nu}$ and $R_{\mu\nu\alpha\beta}$ are the Ricci scalar, Ricci and Riemann tensors correspondingly, $c_i$, $\alpha$, $\beta$ and $\gamma$ are numerical coefficients \cite{litlink1}. The BH solution for the action (\ref{act00}) was obtained and has the form (in $(-, +, + , +)$ signature):  
\begin{equation}\label{eq1} 
ds^2 =  - f_{t} dt^2 + f_{r} dr^2 + r^2 d\Omega^2 ,
\end{equation}
where 
\begin{eqnarray*}
f_r  & \simeq & \left(1 - \frac{2G_{n}M}{r}\right)^{-1} - \frac{\hat{\beta}\hbar{G_{n}^2M}}{r^3} + O(G_n^3) , \\ 
f_t & \simeq & \left(1 - \frac{2G_{n}M}{r}\right) - \frac{\hat{\alpha}\hbar G_{n}^2M}{r^3} + O(G_n^3). 
\end{eqnarray*}
The values $\hat{\alpha}$ and $\hat{\beta}$ are the linear combinations of gauge coefficients from Table 1 in \cite{litlink1}, $M$ is the BH mass and $G_n$ is the effective gravitational constant. 

Here it is necessary to emphasis few important items. The first one is that the structure of non-local actions in different theories could have the same structure. Therefore their BH solutions will also have the same structure. The most developed BH solution is Eq (\ref{eq1}) where one can see the combinations like $G_n M$. This means that the BH mass $M$ is followed by the quantum coefficient $G_n$. Taking into account the real mass of Sgr A* it means the difference of $10^{44}$ between these two values therefore the influence of non-local part is negligible. Anyway we suppose that to proceed the shadow modelling till the finish is interesting from 2 items. One of them is to show the model-independent approach for quantum gravity effects accounting. The other one is to demonstrate that if the non-local action would have the same form (but with coefficients not from the Planckian range) one could apply the suggested scheme to make the theoretical predictions for BH shadows more accurate without adding new fields and/or curvature expansions. 

So as one can see from Eq. (\ref{eq1}) the obtained solution is the spherically-symmetric BH. Following the widely discussed idea that the study of BH shadows image could give an additional information of the extended gravity structure \cite{c04,c05} it seems interesting to use the last results of the Event Horizon Telescope (EHT) \cite{c06}. Here it is necessary to note that the simple estimation of the discussed method application to describe the Universe accelerated expansion (to calculate the turnaround radius \cite{c03b}) gives the negative result by the same reason discussed above. 

Further both objects studied by the EHT represent the spinning BHs. Meanwhile the BH solution established in \cite{litlink1} is non-rotating BH. So to increase the accuracy of theoretical predictions one has to generate a Kerr-like solution from the existing Schwarschild one. At the next step one has to use the Kerr-like metrics for the theoretical modeling. 

To fulfill the suggested program the paper structure is as follows. The Section \ref{s2} is devoted to the Kerr-like BH solution generation, in Section \ref{s3} we discuss the generic items on spinning BH images and the corresponding modelling ideas, The Section \ref{s4} contains the results of our BH shadows modelling in the hypothetical case $G_n\approx M$, Section \ref{s5} is devoted to the comparing of our results with EHT ones and the Section \ref{s6} contains the discussion and our conclusions.     

\section{Adding rotation}\label{s2}

\subsection{Newman-Janis method application}

To obtain a rotating solution the Newman-Janis algorithm could be applied \cite{litlink2,c05}. This algorithm treats a rotating solution as the generation of non-rotating one \cite{litlink3}. According to the algorithm at the first step one has to establish the initial non-rotating metrics in Eddington-Finkelstein coordinates $(u, r, \theta, \phi)$ using the transformation rule:
\begin{align}\label{EF}
dt = \sqrt{\frac{f_{r}}{f_{t}}}du +dr.
\end{align}
Therefore the metrics (\ref{eq1}) takes the form:
\begin{align}\label{EF1}
ds^2 =  - f_{t} du^2 -2 \sqrt{f_{r}f_t}dudr + r^2 d\Omega^2 .
\end{align}
Further we introduce the complex veirbein $e^a=(l^{\mu}, n^{\mu}, m^{\mu}, m^{*\mu}$) with the conditions:
 \begin{equation}\label{flatm}
    \eta^{ab} =
           \left(
    \begin{array}{cccc}
       0 & -1& 0& 0 \\
        -1& 0& 0&  0\\
        0& 0& 0&  1\\
        0& 0& 1& 0
    \end{array}  
    \right),
\end{equation}
and therefore:
\begin{eqnarray*}
l^\mu & = & \delta^{\mu}_{r}, \\
n^\mu & = & \sqrt{\frac{1}{f_rf_t}}\delta^{\mu}_{u} - \frac{1}{2f_r}\delta^{\mu}_{r} \\
m^\mu & = & \sqrt{\frac{1}{2r^2}}\left(\delta^{\mu}_{\theta} + \frac{i}{\sin\theta}\delta^{\mu}_{\phi}\right) \\
m^{*\mu} & = & \sqrt{\frac{1}{2r^2}}\left(\delta^{\mu}_{\theta} - \frac{i}{\sin\theta}\delta^{\mu}_{\phi}\right)
\end{eqnarray*}
To generate a rotation one has to include a complex veirbein transformation in the form:
\begin{eqnarray*}
r \rightarrow r^{'} & = & r - ia \cos{\theta}, \\
u \rightarrow u^{'} & = & u + ia \cos{\theta}, 
\end{eqnarray*}
where $a$ is angular acceleration. After the transformation the functions ${f_t}$, ${f_r}$ and the squared radial coordinate $r^2$ take the form:
\begin{align}\label{f1}
&f_r \rightarrow \tilde F_{r}(r,\theta,a), \\
&\lim_{a\to0}\tilde F_{r}(r,\theta,a) = f_r, \\
&f_t \rightarrow \tilde F_{t}(r,\theta,a) , \\ \label{f2}
&\lim_{a\to0}\tilde F_{t}(r,\theta,a) = f_t, \\
&r^2 \rightarrow \rho^2 = r^2 + a^2\cos^2{\theta}. 
\end{align}
Following \cite{litlink4} it is necessary to note that the transformations   (\ref{f1} - \ref{f2}) are not unique and additional conditions are required. The most convenient choice is $g_{rt} = g_{r\phi} = 0$. Therefore the renewed veirbein takes the form: 
\begin{eqnarray}
{l^{\mu}}' & = & \delta^{\mu}_{r'}, \label{tet2} \\  
{n{^\mu}}' & = & \sqrt{\frac{1}{\tilde F_r \tilde F_t}}\delta^{\mu}_{u'} - \frac{1}{2f_r}\delta^{\mu}_{r'}, \label{tet3} \\ 
{m{^\mu}}' & = & \sqrt{\frac{1}{2\rho^2}}\left(\delta^{\mu}_{\theta} + ia\sin\theta(\delta^{\mu}_{u'} -\delta^{\mu}_{r'}) + \frac{i}{\sin\theta}\delta^{\mu}_{\phi}\right). \label{tet4}
\end{eqnarray}
Using equations (\ref{flatm}) and (\ref{tet2})-(\ref{tet4}) one obtains the BH metrics where the rotation is now included:  
\begin{eqnarray}
ds^2 = & - & \tilde F_t du^2 - 2 \sqrt{\tilde F_r \tilde F_t} du dr + \rho^2d\theta^2 \nonumber \\
   & - & 2a \sin^2 \theta (\sqrt{\tilde F_r \tilde F_t} -\tilde F_t) du d\phi \nonumber \\
   &+ & 2a \sin^2\theta (\sqrt{\tilde F_r \tilde F_t}dr d\phi \nonumber \\ & + &\sin^2\theta\left(\rho^2 + a^2 \sin^2 \theta (2\sqrt{\tilde F_r \tilde F_t} -\tilde F_t)\right)d\phi^2. 
\end{eqnarray}
Finally the transverse transformation in the form  
\begin{eqnarray}
du & = &dt + \chi_1(r) dr, \nonumber  \\
d\phi &= & d\varphi + \chi_2(r) dr
\end{eqnarray}
is applied. Following \cite{litlink4} we take $\chi_1(r)$ and $\chi_2(r)$ as:
\begin{align}
\chi_1 &= -\frac{f_r(\omega+a^2)}{r^2+a^2 f_r}, \quad  \\
\chi_2 &= -\frac{f_r a}{r^2+a^2f_r},\\\label{omega}
\omega &= r^{2} \sqrt{ \frac{1}{f_r f_t} }.
\end{align}
So the final form of Kerr-like metrics for the discussed model is:
\begin{eqnarray}\label{eq30}
g^{tt} & = & -\frac{1}{\rho^2}\left[\frac{(\omega+a^2)^2}{(f_r^{-1} r^2+a^2)}-a^2 \sin^2\theta \right],\nonumber \\ 
g^{t\phi} & = & -\frac{1}{\rho^2}\left[\frac{(\omega+a^2)a}{(f_r^{-1} r^2 + a^2)}-a\right],\nonumber \\
g^{\phi \phi} &= & -\frac{1}{\rho^2}\left[\frac{a^2}{ (f_r^{-1}r^2+a^2)}-\frac{1}{ \sin^2 \theta}\right],\nonumber \\
g^{\theta \theta} & = & \frac{1}{\rho^2},\quad g^{rr}=\frac{f_r^{-1} r^2 +a^2}{\rho^2}.
\end{eqnarray}
Here 
\begin{eqnarray*}
\rho^2 & = & r^2 + a^2 \cos^2\theta, \\
f_{s} & = & 1 - \frac{2MG_{n}}{r}, \\
f_{ex}& = & \frac{\hbar G_{n}^2M}{r^3}, \\
\omega & = & r^2 \left( 1 + \frac{(\hat{\alpha} +\hat{\beta})f_{ex}}{2} \right).
\end{eqnarray*}

\subsection{The Hamilton-Jacobi equation}

To be able to derive the photons trajectories around the spinning BH the form of $S_r(r)$ and $S_\theta(\theta)$ from Hamilton-Jacobi equation is required. For null geodesics: 
\begin{equation}\label{eq:HJE}
    g^{\mu \nu} \frac{\partial S}{\partial x^\mu} \frac{\partial S}{\partial x^\nu} =0.
\end{equation}
As the obtained metric has no dependence upon $t$ and $\phi$ therefore 2 conserved quantities occur. They are: $E=-p_t$ and $L_z=p_\phi$ (photon's energy and angular momentum respect symmetry axis). Therefore to divide the variables one has to look for the solution of equation (\ref{eq:HJE}) in the form:
\begin{equation}
    S=-Et+L_z\phi+S_r(r)+S_\theta(\theta).
\end{equation}
Next it is necessary to conclude that the variables in eq. (\ref{eq:HJE}) for $p_r$ and $p_\theta$ can be divided as:
\begin{align}
\frac{\rho^4 (\dot{r})^2}{E^2}&=\mathcal{R}(r),\\
\frac{\rho^4 (\dot{\theta})^2}{E^2}&=\Theta(\theta),
\end{align}
where 
\begin{align}
    \label{eq:R}
    \mathcal{R}(r)&=\left(\omega + a^2 - a\lambda\right)^2-(f_r^{-1} r^2 +a^2)\left[\eta+\left(a-\lambda\right)^2\right],\nonumber \\
    \Theta(\theta)&=\eta+\cos^2 \theta \left(a^2-\frac{\lambda}{ \sin^2 \theta}\right).
\end{align}
Here $\eta=\frac{Q}{E^2}$, $\lambda=\frac{L_z}{E}$ and $Q$ is Carter's constant.

To calculate the equation for the circular photon orbit (only such photons could reach an remote observer) 
\begin{align}
    \label{eq:R=0}\mathcal{R}&=0,\\
    \label{eq:dR/dr=0}\frac{d\mathcal{R}}{dr}&=0.
\end{align}
Substituting eq. (\ref{eq:R}) to (\ref{eq:R=0}) and (\ref{eq:dR/dr=0}) one finds the solutions for $\lambda$ and $\eta$ in the form:
\begin{eqnarray}
\lambda & = & \frac{\omega+a^2}{a}-\frac{2\omega'}{a}\frac{(f_r^{-1}r^2+a^2)}{(f_r^{-1}r^2)'},\label{eq:lambda} \\
\eta & = & \frac{4(f_r^{-1}r^2+a^2)}{{(f_r^{-1}r^2)'}^2}{\omega'}^2-\frac{1}{a^2}\left[\omega - \frac{2(f_r^{-1}r^2+a^2)}{(f_r^{-1}r^2)'}\omega'\right]^2, \nonumber 
\end{eqnarray}
where strokes denote the derivatives respect to $r$. Considering the plane normal to the direction to the remote observer the shadow coordinates cab be written as follows 
\begin{align}
    x'&=-\frac{\lambda}{\sin\theta_0}, \label{eq45}\\
    y'&=\pm \sqrt{\eta+a^2 \cos^2\theta_0 - \frac{\lambda^2}{\tan^2\theta_0}}, \label{eq46}
\end{align}
where $\theta_0$ is solid angle between the BH rotation plane and the axis to observer.

\section{The modelling of BH shadow: how to include the rotation}\label{s3}

With help of Python coding language we proceed the numerical modelling of the BH shadow from the metrics (\ref{eq30}). Using expressions (\ref{eq45}) and (\ref{eq46}) coordinates on the picture plane $X$ and $Y$ на were calculated. The metrics (\ref{eq30}) was applied with the different values of rotation characteristics $a$ and the coefficients $\alpha$ and $\beta$.

\begin{figure}
\begin{center}
\includegraphics[scale=1.5]{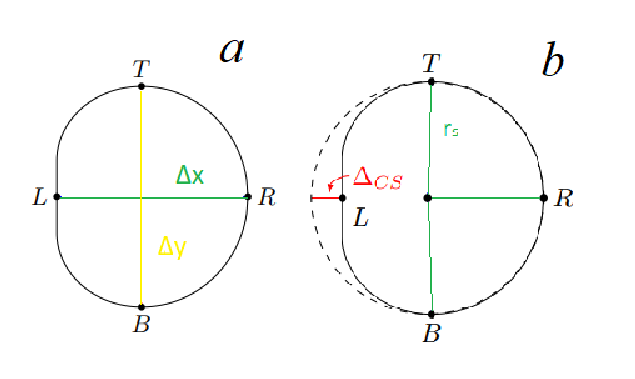}
\end{center}
\caption{The shadows limiting values $R$, $L$, $T$ and $B$ against shadow diameters $\Delta x$ and $\Delta y$ (a). The observable $r_s$ and $\Delta_cs$, obtained from the circle passing $T$, $R$ and $B$ (b).}
\label{diam_delt}
\end{figure}

As it was demonstrated earlier \cite{litlink5} the shadow from spinning BH has some particular properties:
\begin{enumerate}

\item {\bf Horizontal shift}
The shadow shift along $x$ axis could be calculated with the help of the expression:
\begin{equation}\label{bhs1} 
D = \frac{x_{min}+x_{max}}{2} ,
\end{equation}
where $x_{min}$ and $x_{max}$ are the minimal and maximal $x$ values respectively. 

\item {\bf Asymmetry}
When the values of spinning parameter $a$ increase and become rather large the shadow asymmetry appears \cite{c05} so that the horizontal radius value becomes less than the vertical one which does not change. Therefore the horizontal shadow size becomes a measure of asymmetry:
\begin{equation}\label{bhs2} 
\Delta x = x_{max} - x_{min}.
\end{equation}

\item {\bf Diameters}
Let's treat the horizontal diameter as:
\begin{equation}\label{bhs3} 
\Delta x = x_{max} - x_{min} = x_R - x_L ,
\end{equation}
where $L$ and $R$ are left and right shadow shape points. One can define the vertical diameter in the same manner as:
\begin{equation}\label{bhs4} 
\Delta y = y_{max} - y_{min} = y_T - y_B = 2y_T ,
\end{equation}
where $B$ and $T$ are the bottom and top points of the shadow edge. Due to the symmetry of the shadow: $y_B = y_T$, and Fig.~\ref{diam_delt}(a) shows the relationship between $\Delta y$, $\Delta x$ and points $R, L, T$ and $ B$.

\item {\bf Circular approximation}
As the shadow has the quasi-circular form it is convenient to treat the points $T$, $R$ and $B$ lying on the circular \cite{litlink6}. Hence the shadow radius $r_s$ becomes the first observational value. As the second observational value one can use the distortion parameter:
\begin{equation}\label{bhs5} 
\delta_{cs} = \Delta_{cs}/r_s ,
\end{equation}
where $\Delta_{cs}$ is the distance from the circular to the point  $L$ on the shadow (See Fig.~\ref{diam_delt}(b)).
\end{enumerate}

\section{The modelling of BH shadow for the space-time (\ref{eq30})}\label{s4}

After few preliminary notes one can start to calculate the shadow dependence upon $\alpha$ and $\beta$. Applying gravitational corrections to the stable star metric satisfying Tolmen-Oppenheimer-Volkov equation \cite{litlink7} we introduce new variables $\alpha=\hat{\alpha}$ и $\beta=\hat{\beta}$ that are model-independent. 

\begin{figure}
\begin{center}
\includegraphics[width=\linewidth]{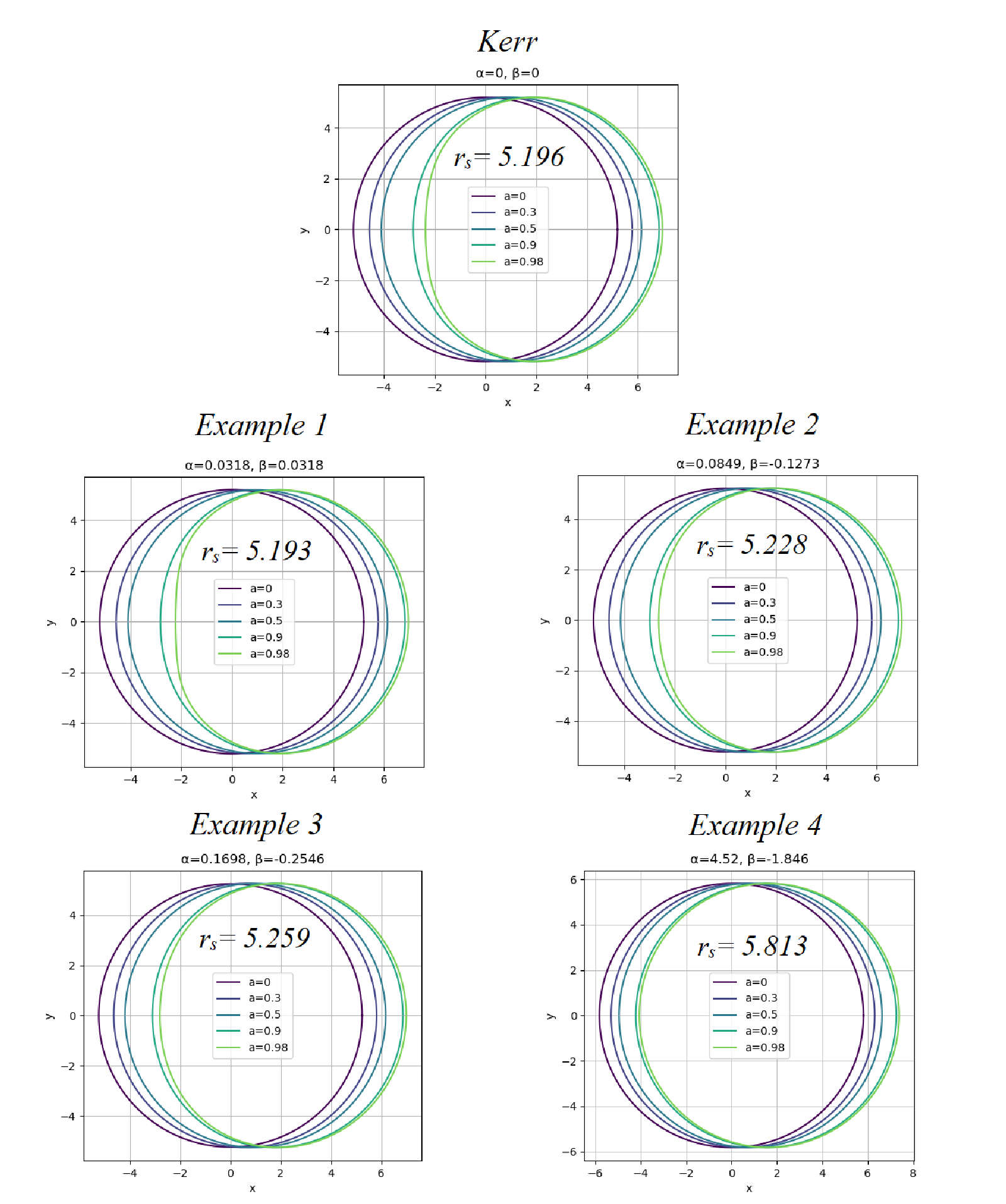}
\end{center}
\caption{The shadow profile for different rotation parameters $a$ for Kerr-like metric and its generalisations when the inclination angle of the rotation plane is $\theta_0=\frac{\pi}{2}$.}
\label{r+r2}
\end{figure}

Here it is necessary to point out that we use the coefficient values from \cite{litlink1} as examples. So we establish the shadow shapes for $M=1$\footnote{Because in the real case $M=10^{44}$ the effect is vanishing as it was pointed out in the Introduction} and different $a$ values for Kerr metric and its extensions defined at \cite{litlink7} (in the scalar field $\xi=1/3$) on Fig.~\ref{r+r2}. The angle of plane of rotation is equal to $\theta_0=\frac{\pi}{2}$. Note the 2 main particularities. Firstly the shadow shape shifts from rotation axis with the increasing of $a$. Secondary the shadow becomes asymmetric along $x$ axis for big values of $a$. Both particularities vanish at  $a \rightarrow 0$ when the circular shadow corresponding to Schwarzchield case reduces. Also note that when $\theta_0=\pi/2$ the shadow size does not change depending upon rotation (because the vertical diameter remains the same).

For different field types we obtained the following effective shadow sizes $r_s$ (Table 1):
\begin{table}[h]
\label{tabb1}
\centering
\begin{tabular}{|c|c|c|c|}
\hline
Solution type & $\alpha$ & $\beta$ & $r_s$ \\
\hline
Kerr & 0 & 0 & 5.196 \\
Example 1 & 0.0318 & 0.0318 & 5.193 \\
Example 2 & 0.0849 & -0.1273 & 5.228 \\
Example 3 & 0.1698 & -0.2546 & 5.259 \\
Example 4 & 4.52 & -1.846 & 5.813 \\
\hline
\end{tabular}
\end{table}

As EHT constrained the shadow size as ($4.3M<r_s<5.3M$)  \cite{SgrA} one conditionally could neglect the last line in the table 1.

\begin{figure}
\begin{center}
\includegraphics[width=\linewidth]{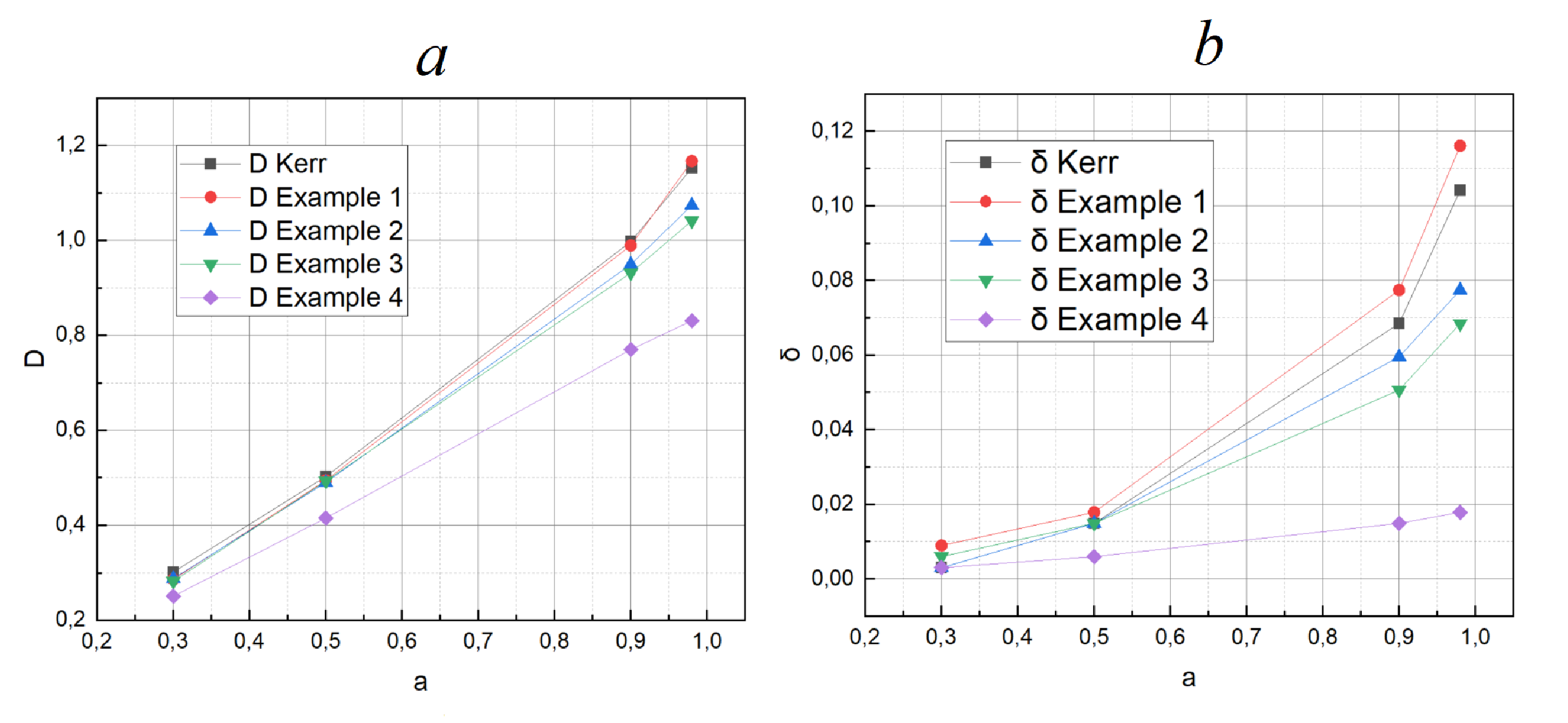}
\end{center}
\caption{The dependence of the shift $D$ (a) and distortion parameter $\delta$ (b) against rotation acceleration $a$ for quasi-Kerr metric and for different fields added when the inclination angle of the rotation plane is $\theta_0=\frac{\pi}{2}$.}
\label{D_delta}
\end{figure}

Now we concentrate on BH images study. We are interesting in the shift $D$ and and distortion parameter $\delta$. On Fig.~\ref{D_delta}(a) we show the dependence of $D$ against rotation parameter $a$ in all discussed case. While $\alpha$ and $\beta$ increase the shift becomes smaller. The only exception is the case of near extreme rotation with $a=0.98$ and the value for the Example 1. We suppose that the reason is that in the discussed case $\beta>0$ and in other ones it is negative. In the case of big $\alpha$ and $\beta$ values the shift has the linear dependence against $a$. Unfortunately, in practice it is difficult to extract the value of this parameter as there is no information on the coordinate origin. The distortion parameter $\delta$ is shown on the Fig.~\ref{D_delta}(b)). The most difference from spherical form occurs in Example 1. At Example 4 case in $\alpha$ and $\beta$ have rather big absolute values the shadow remains spherically-symmetric even in the big $a$ case.

\section{Constraints from EHT results on Sgr A*}\label{s5}

\begin{figure}
\begin{center}
\includegraphics[width=\linewidth]{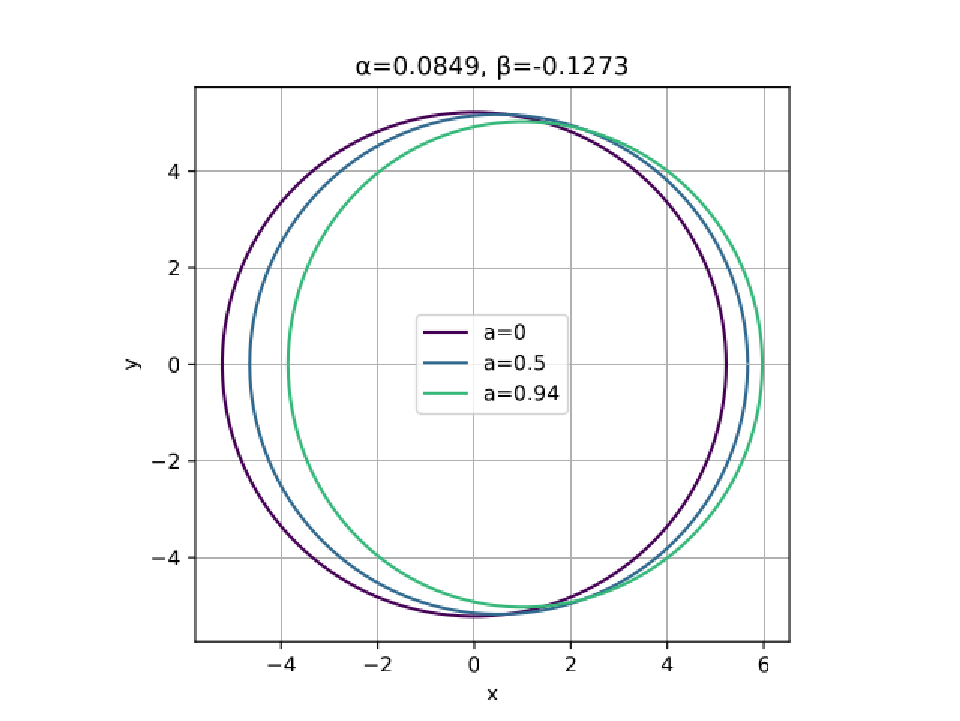}
\end{center}
\caption{The BH shadow profile against rotation acceleration $a$ for example 2 field  when the inclination angle of the rotation plane is $\theta_0=\frac{\pi}{6}$ (Sgr A*).}
\label{SgrA}
\end{figure}

\begin{figure}
\begin{center}
\includegraphics[width=\linewidth]{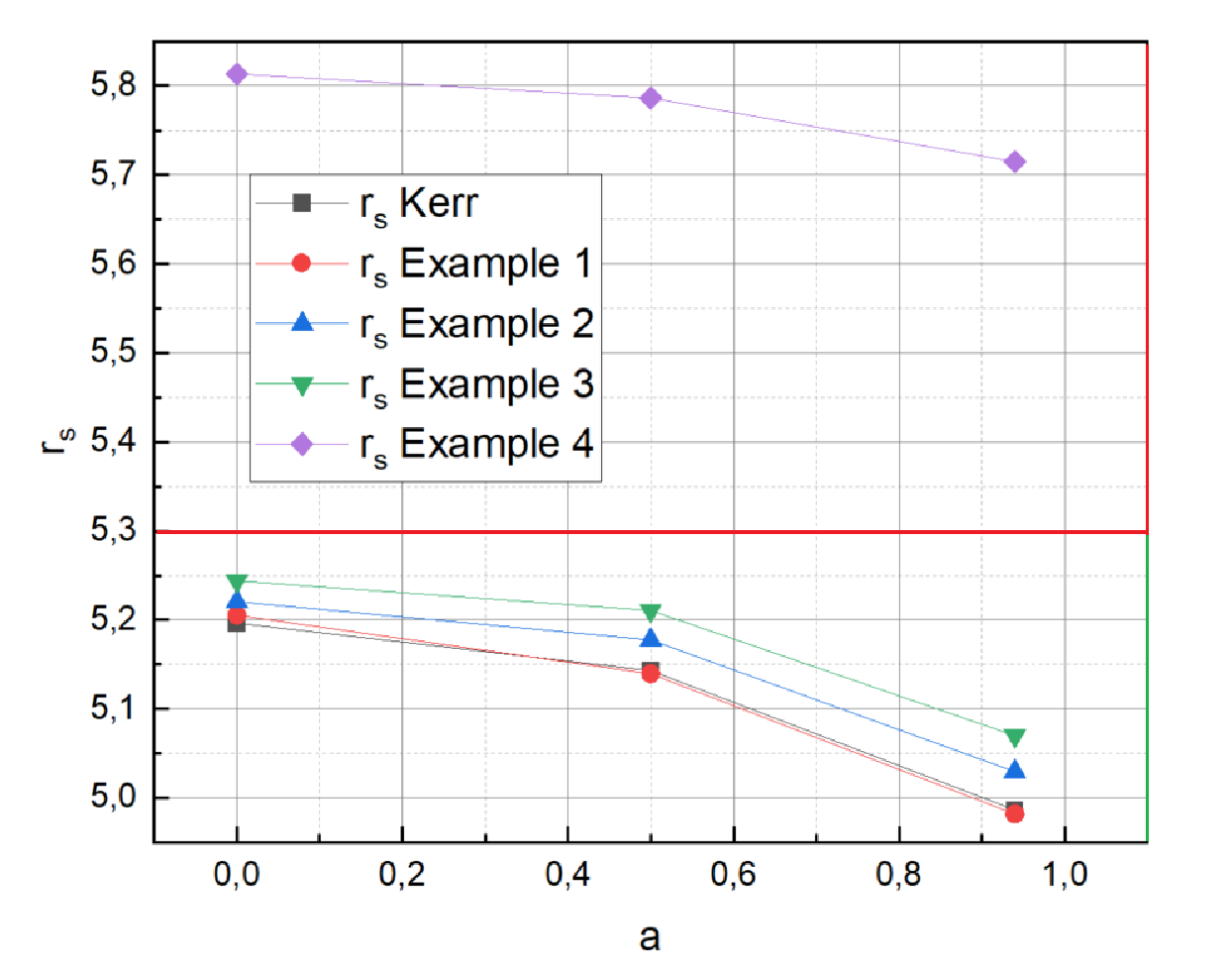}
\end{center}
\caption{The dependence of BH shadow size $r_s$ against rotation acceleration $a$ for Kerr-like metric and with the additional fields added when the inclination angle of the rotation plane is $\theta_0=\frac{\pi}{6}$ (Sgr A*). The red line (more than 5.3) denotes the region excluded by EHT results on Sgr A*, the green line (less than 5.3) denotes the region allowed by EHT results on Sgr A*.}
\label{rs_sgra}
\end{figure}

\begin{figure}
\begin{center}
\includegraphics[width=\linewidth]{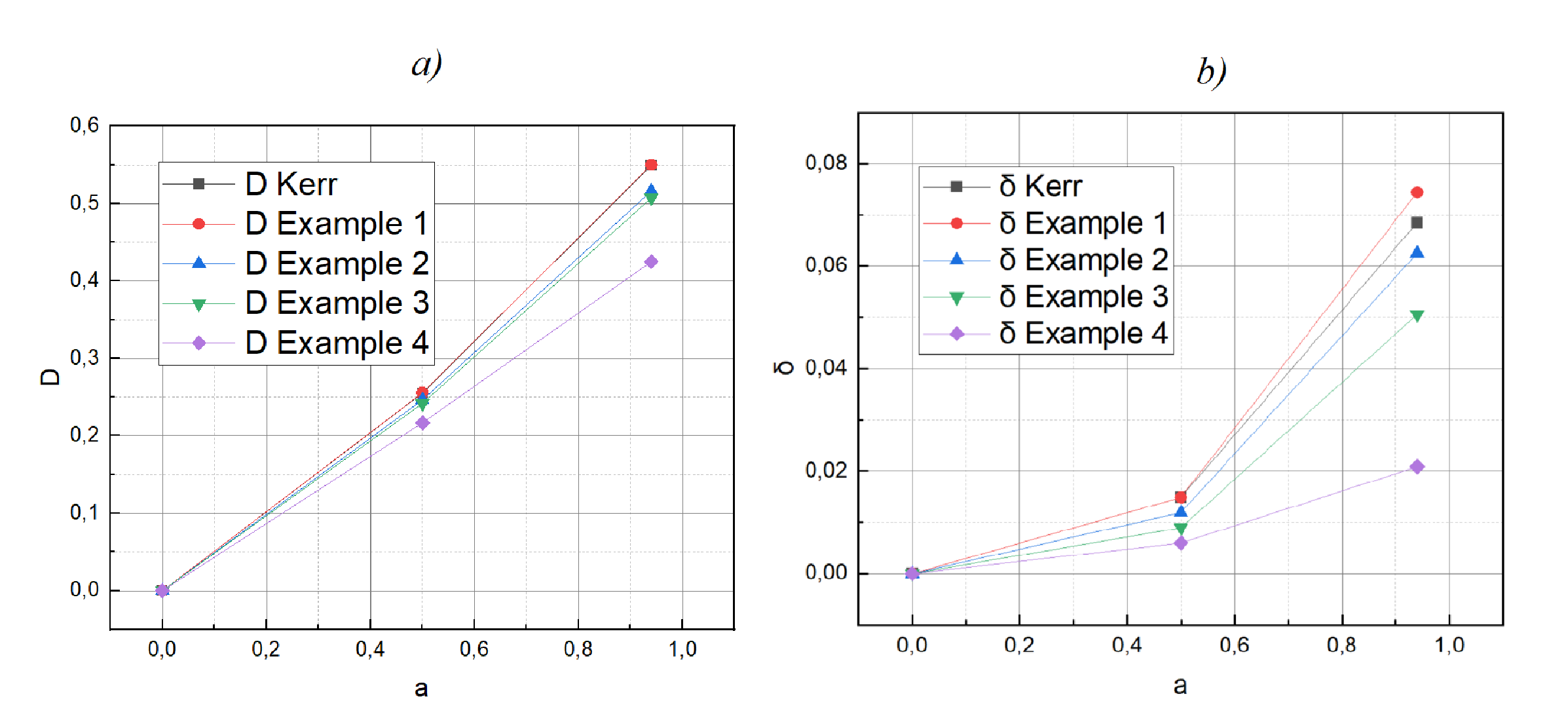}
\end{center}
\caption{The dependence of the shift $D$ (a) and distortion $\delta$ (b) against rotation acceleration $a$ for Kerr-like metrics and with the additional fields added when the inclination angle of the rotation plane is $\theta_0=\frac{\pi}{6}$ (Sgr A*).}
\label{D_delta_sgra}
\end{figure}

Now we have all information necessary to compare our results with EHT images for  Sgr A*. EHT claims that the most probable values of $a$ are equal to $0.5$ and to $0.94$ \cite{SgrA}. This assumption appeared to be possible as Sgr A* is situated in our Galaxy and the orbits of surrounding stars were observed. Fig.~\ref{SgrA} demonstrates the profile of BH shadow for Sgr A* from EHT data (rotation plane inclination is equal to $\theta_0=\pi/6$, $a$ is equal to $0.5$ и $0.94$). For comparison the case $a=0$ is also presented. From fig.~\ref{SgrA} one can conclude that for the given angle value the shape distortion is small but the shadow size changes. We start from the shift of shadow size (fig.~\ref{rs_sgra}). In contrast with the case where $\theta_0=\pi/2$ if $\theta_0=\pi/6$ the shadow size depends against $a$. From the plot it is seen that EHT constraint line passes all the fields (green region) except example 4 one (reg region). Concerning to the shift  $D$ (fig.~\ref{D_delta_sgra}(a)) one concludes it becomes less than in the previous case from previous paragraph. As for the distortion parameter $\delta$ (fig.~\ref{D_delta_sgra}(b)): it has a maximum at $a=0.94$ about 5-8\% (except neglected example 4). At $a=0.5$ the distortion is equal to 1.5\%. 

\section{Discussion and Conclusions}\label{s6}

Using the Newman-Janis algorithm we obtained spinning solution for the quantum gravity model with the action (\ref{act00}). We proceed the modelling of BH shadow for the rotating metrics for the pure quasi-Kerr case and taking into account the additional fields in the limit $G_n\approx M$. For a more visual representation we take $\theta_0=90^\circ$ as at such a case a shadow obtains maximal distortions during the fast rotation. We show that for fast rotation (when $a$ tends to $1$) for all cases except example 4 one the BH shadow is deformed. For the pure Kerr metrics and example 1 this deformation is equal to 10-11\%, for the example 2 and example 3 is drops to 5-8\%. Here it is necessary to point out that less accuracy is required to fix this deformation than to fix the shadow size. This occurs because of 2nd and 3rd order corrections as it was shown earlier \cite{c04}. In the discussed approximation $G_n \approx M$ corrections and a rotation contribute opposite one to the other and, therefore, could compensate each other. So in the future to follow the increasing of the experimental accuracy and therefore to reproduce new results theoretically one could take into account non-local terms (if compatible with BH mass) instead of introducing of new fields and/or curvature expansions.

EHT results \cite{SgrA} show that the most probable tilt angle for Sgr A* is $\theta_0=30^\circ$ and the most probable values of $a$ are equal to $0.5$ and $0.94$. We demonstrated that in this case (in the hypothetical case $G_n\approx M$) the shadow deformation is not big. At $a=0.94$ deformation appears to be 5-8\% (except Example 4), when $a=0.5$ deformation drops to 1.5\%. Therefore after increasing the accuracy the rotation characteristics of Sgr A* could be extracted better. Next, for a given inclination of the rotation plane the shadow size depends upon $a$. This fact could also help to establish BH shadow properties in future. 

Finally the algorithm of non-local gravitational effects accounting in BH shadow modelling was suggested. This algorithm is independent on the ultra-violet structure of complete theory of quantum gravity and could be extended to other non-local theories. 

\section{Acknowledgements}

The work was supported by the Russian Science Foundation via grant No. 23-22-00073.

\newpage
\onecolumn

\vskip40mm

\newpage

\newpage

\newpage

\end{document}